\documentclass[12pt]{iopart}
\usepackage{graphicx}
\usepackage{cite}
\usepackage{color}
\usepackage{amsfonts}

\begin{document}

\title[Nonequilibrium phase transition in a spreading process on a timeline]
      {Nonequilibrium phase transition in a spreading process on a timeline}
\author{Andre C. Barato and Haye Hinrichsen}
\address{Universit\"at W\"urzburg\\
	 Fakult\"at f\"ur Physik und Astronomie\\
         D-97074 W\"urzburg, Germany}
\ead{barato@physik.uni-wuerzburg.de}

\def\d{\mathrm{d}}
\definecolor{darkbrown}{rgb}{0.4,0.1,0.1}
\def\noiseamp{c}

\begin{abstract}
We consider a nonequilibrium process on a timeline with discrete sites which evolves by a non-Markovian update rule in such a way that an active site at time $t$ activates one or several sites in the future at time $t+\Delta t$. The time intervals $\Delta t$ are distributed algebraically as $(\Delta t)^{-1-\kappa}$, where $0<\kappa<1$ is a control paramter. Depending on the activation rate, the system display a nonequilibrium phase transition which may be interpreted as directed percolation transition driven by temporal L{\'e}vy flights in the limit of zero space dimensions. The critical properties are investigated by extensive numerical simulations and compared with field-theoretic predictions.
\end{abstract}

\submitto{Journal of Statistical Mechanics: Theory and Experiment\\
\hspace*{24mm} (SigmaPhi2008 conference proceedings)}
\maketitle


\def\xvec{{\vec x}}

\def\d{{\rm d}}
\def\te{{t_{\scriptscriptstyle \hspace{-0.3mm} e}}}
\def\tf{{t_{\scriptscriptstyle \hspace{-0.3mm} f}}}
\def\Ps{{P_{\scriptscriptstyle \hspace{-0.3mm} s}}}
\def\Pe{{P_{\scriptscriptstyle \hspace{-0.3mm} e}}}

\newpage

\tableofcontents

\newpage
\vspace{25mm}
\parskip 2mm 
\pagestyle{plain}

\section{Introduction}

\begin{figure}[t]
\begin{flushright}
\includegraphics[width=150mm]{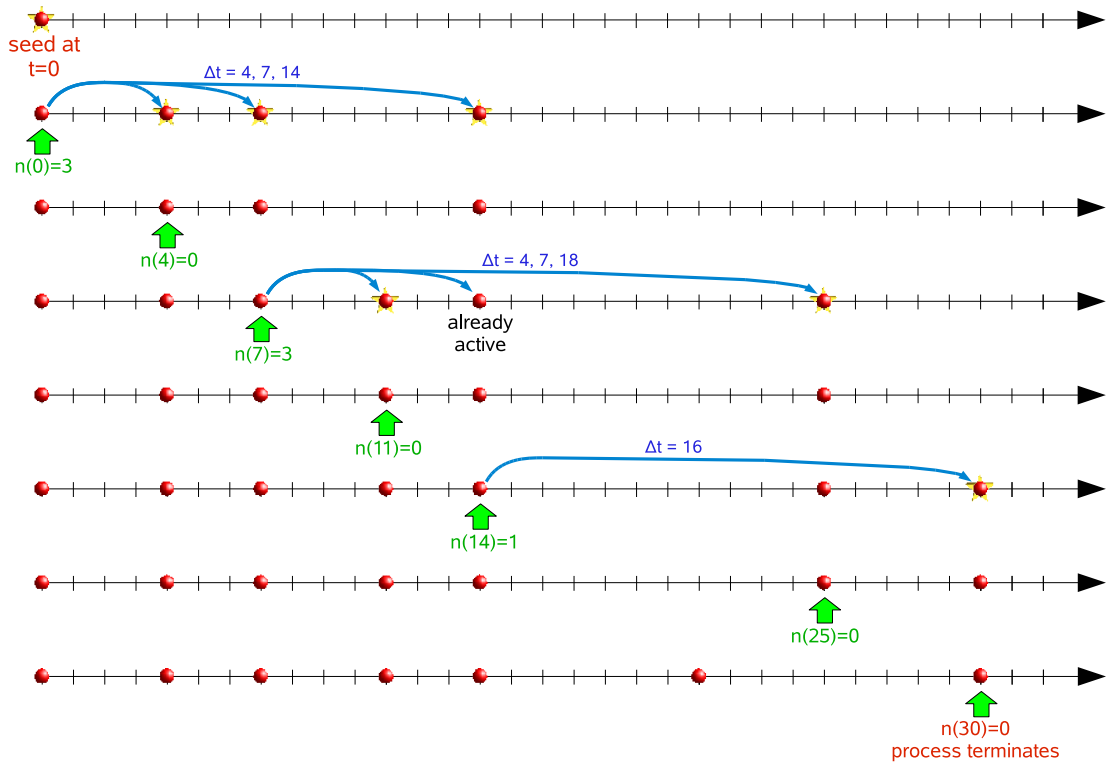}
\end{flushright}
\vspace{-5mm}
\caption{Timeline spreading process (TSP): Example of a temporal evolution starting with a single active site at $t=0$. Active sites are updated from left to right, as marked by the green arrows. The blue arrows indicate how activity spreads by L{\'e}vy flights over randomly chosen displacements $\Delta t$.}
\label{fig1}
\end{figure}

In the present paper we consider a class of probabilistic spreading processes on a timeline. As sketched in Fig.~\ref{fig1}, the timeline consists of discrete sites $t \in \mathbb{Z}$ which can be either active or inactive. After specifying a certain initial configuration of active and inactive sites along this timeline, the process evolves dynamically by subsequent updates of the lattice site according to the following probabilistic rules:
\begin{itemize}
\item If the updated site at time $t_u$ is active, it attempts to active $n(t_u)$ lattice sites in the future, where $n(t_u)=0,1,2,\ldots$ is randomly selected from a given distribution with a finite average $\bar n$. 
\item For each of these attempts a random time interval $\Delta t=1,2,\ldots$ is drawn from another probability distribution $P(\Delta t)$. If the target site at time $t_u+\Delta t$ is still inactive it will be activated, otherwise nothing happens.
\end{itemize}
In the following we are interested in the special case where the time intervals are asymptotically distributed by a power law
\begin{equation}
\label{Distribution}
P(\Delta t) \sim (\Delta t)^{-1-\kappa} \,,
\end{equation}
where $0<\kappa<1$ is a control parameter. In this case the process can be considered as a model for the spreading of activity on a timeline by means of temporal L{\'e}vy flights~\cite{ShlesingerEtAl95a,Fogedby94}.

Remarkably, the timeline spreading process (TSP) shown in Fig.~\ref{fig1} displays a dynamical non-equilibrium phase transition when the parameter $\bar n$ is varied. For example, if $\bar n$ is sufficiently small the process starting with a single active site at $t=0$ will terminate after some time, while for large $\bar n$ the process may survive forever, producing an asymptotically constant density of active sites. As will be shown below, the two regimes of survival end extinction are separated by a well-defined critical threshold $\bar n_c$, where the system undergoes a continuous phase transition. Despite the simplicity of the model this transition turns out to be characterized by a surprisingly non-trivial critical behavior~\cite{DeloubriereWijland02,BaratoHinrichsen08a}.

The other interesting aspect of the TSP is the non-Markovian nature of its dynamics. Since activity may spread over arbitrarily long time intervals $\Delta t$, the actual state of an updated site depends not only on the previous time step but rather on the entire history of the process. Therefore, the initial condition is not determined by the state of the system at $t=0$ alone, instead one has to specify the complete configuration of active sites along the entire timeline. In fact, the TSP is probably the simplest model which allows one to study non-Markovian features in the context of continuous phase transitions far from equilibrium. 

The transition in the TSP belongs to the category of so-called absorbing phase transitions~\cite{Hinrichsen00,Odor04,Lubeck04,Odor08a,HenkelEtAl08a} because it may be interpreted as a transition from a fluctuating active state into a frozen inactive state where the process terminates. Such transitions are associated with certain universality classes, the most prominent one being the universality class of directed percolation (DP)~\cite{Kinzel85}, which plays a similar role as the Ising class in equilibrium statistical mechanics. It is represented e.g. by the contact process~\cite{Liggett85} which is a toy model for epidemics where activity spreads to nearest neighbors on a $d$-dimensional lattice by means of a Markovian update rule. Although this model is easy to define it could be realized experimentally only one year ago by Takeuchi \etal~\cite{TakeuchiEtAl07}. Recently DP was generalized to include long-range interactions by incorporating spatial~\cite{JanssenEtAl99,HinrichsenHoward99,Janssen08a} and temporal~\cite{JimenezDalmaroni06} L{\'e}vy flights as well as a combination of both~\cite{AdamekEtAl05} (for a review see~\cite{Hinrichsen07}). As we will see, the TSP studied here may be interpreted as the zero-dimensional limit of DP with temporal L{\'e}vy flights because here we have a single site that evolves in time. 

\begin{figure}[t]
\begin{flushright}
\includegraphics[width=150mm]{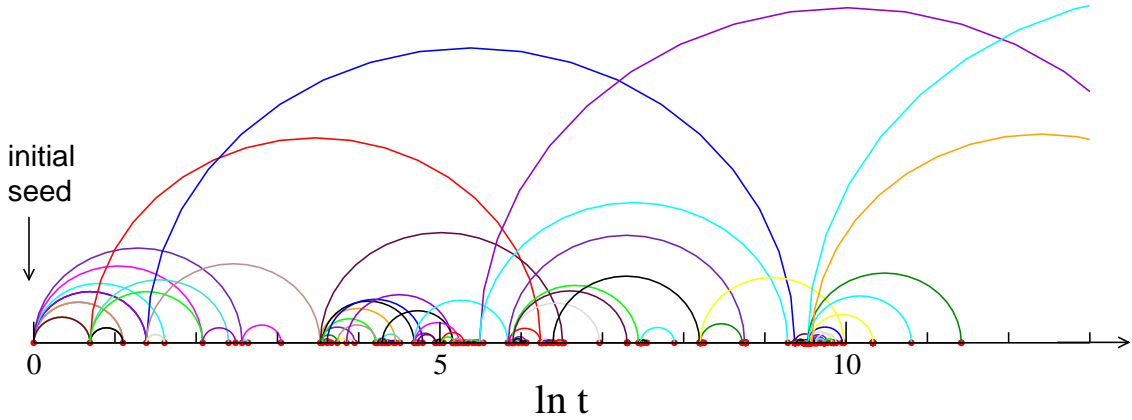}
\end{flushright}
\caption{Typical temporal evolution of the TSP for $\kappa=0.5$ on a logarithmic time scale. The red dots indicate active sites while the half circles illustrate the L{\'e}vy flights directed forward in time. As can be seen, activity occurs in form of intermittent bursts.}
\label{fig2}
\end{figure}

From a broader perspective, the TSP is also of conceptual interest. According to a well-known theorem by Landau, phase transitions in \textit{equilibrium} models with short-range interactions require at least two space dimensions. In the non-equilibrium case, however, phase transitions in models with short-range couplings such as DP are possible in one spatial dimension. The TSP demonstrates that by introducing a non-Markovian (i.e. temporally non-local) dynamics phase transitions are possible even in zero space dimensions.

The paper is organized as follows. In the following section we introduce the specific variant of the TSP studied in this paper and discuss phenomenological properties of the phase transition. In Sect. 3-4 we summarize previous field-theoretic results~\cite{DeloubriereWijland02} which allow one to identify a mean-field and a fluctuation-dominated regime and to compute the critical exponents to one-loop order. In Sect. 5 we confirm these findings by numerical simulations whereas the conjugate field and the special role of the survival probability will be discussed in Sect.~6. The relation to a recently studied boundary-induced phase transition in 1+1 dimensions~\cite{DeloubriereWijland02,BaratoHinrichsen08a} will be discussed in a forthcomming publication.\\ 

\section{The model and its phenomenological properties}

\subsection{Definition and numerical implementation}

There are many possible variants of the TSP which differ mainly in their specific probability distribution for the number of attempted activations $n(t)$ and in the short-range details of the L{\'e}vy distribution $P(\Delta t)$. These variants may have different critical thresholds~$\bar n_c$ but for given $\kappa \in (0,1)$ their critical properties at the transition are expected to be universal. 

The numerical results reported in this paper were obtained by simulating the following variant of the TSP. It is defined on a timeline with discrete sites $s(t)$ at time $t \in \mathbb{Z}$, where the values $s(t)=0,1$ denote inactive and active sites, respectively. For each run one first has to specify the initial configuration by assigning certain values to \textit{all} sites. In most cases we will start with a single seed of activity at $t=0$ by setting $s(t):=\delta_{t,0}$. After specifying the initial state an update loop over all time steps $t_u$ is executed which starts with the minimal $t$ for which $s(t)=1$ and ends when the process terminates, limited by some cutoff time $t_{\rm max}$. Inside this loop the following steps are carried out:
\begin{quote}
\begin{enumerate}
\item If $s(t_u)=0$ go to (v).\\[-3mm]
\item Generate a random number $z \in (0,1)$. If $z>p$ go to (v).\\[-3mm]
\item Generate another random number $y \in (0,1)$ and set $\Delta t:= y^{-1/\kappa}$.\\[-3mm]
\item If $t_u+\Delta t<t_{\rm max}$ activate the target site by setting $s(t_u+\lfloor\Delta t\rfloor):=1$, where $\lfloor.\rfloor$ denotes truncation to an integer, and go back to step (ii).\\[-3mm]
\item Increment $t_u$ and proceed with the next time step.
\end{enumerate}
\end{quote}
This particular variant of the TSP is controlled by the spreading probability $p\in [0,1]$ and the L{\'e}vy exponent $\kappa \in (0,1)$. As can be verified easily, the assignment $\Delta t:= y^{-1/\kappa}$ generates a probability distribution with a lower cutoff at $\Delta t=1$ which reproduces the asymptotic decay postulated in Eq.~(\ref{Distribution}). Moreover, the model is defined in such a way that the number of attempted activations $n(t_u)$ discussed in the previous section occurs with probability $p^{n}(1-p)$, hence $\bar n = p/(1-p)$. Note that in this update scheme repeated activations of the same target site have no effect. 

The above algorithm can be easily implemented on a computer. For example, if {\tt rnd()} returns a random number drawn from a flat distribution between 0 and 1, a minimal C-code for this update procedure would read as follows:
\color{darkbrown}
\begin{quote}
\begin{small}
\begin{verbatim}
const int Tmax=10000;                    // maximal cutoff time;
const double kappa=0.5, p=0.574262;      // control parameters;
int s[Tmax];                             // the timeline;

for (int t=0; t<Tmax; ++t) s[t]=0;       // clear timeline;
s[0]=1;                                  // place initial seed;
for (int tu=0; tu<Tmax; ++tu)            // execute update loop
  if (s[tu]==1)                          // over all active sites
    while (rnd()<p) {                    // repeatedly with prob. p;
      double dt = pow(rnd(),-1/kappa);   // generate a time interval
      if (tu+dt<Tmax) s[tu+floor(dt)]=1; // and activate target site.
      }
\end{verbatim}
\end{small}
\end{quote}
\color{black}
Depending on the initial state, the algorithm can be accelerated significantly by storing the active sites in a dynamically generated list instead of using a static array. The structure of such an optimized code is outlined in the appendix at the end of this paper.

\subsection{Phenomenological properties}

\begin{figure}[t]
\centering\includegraphics[width=130mm]{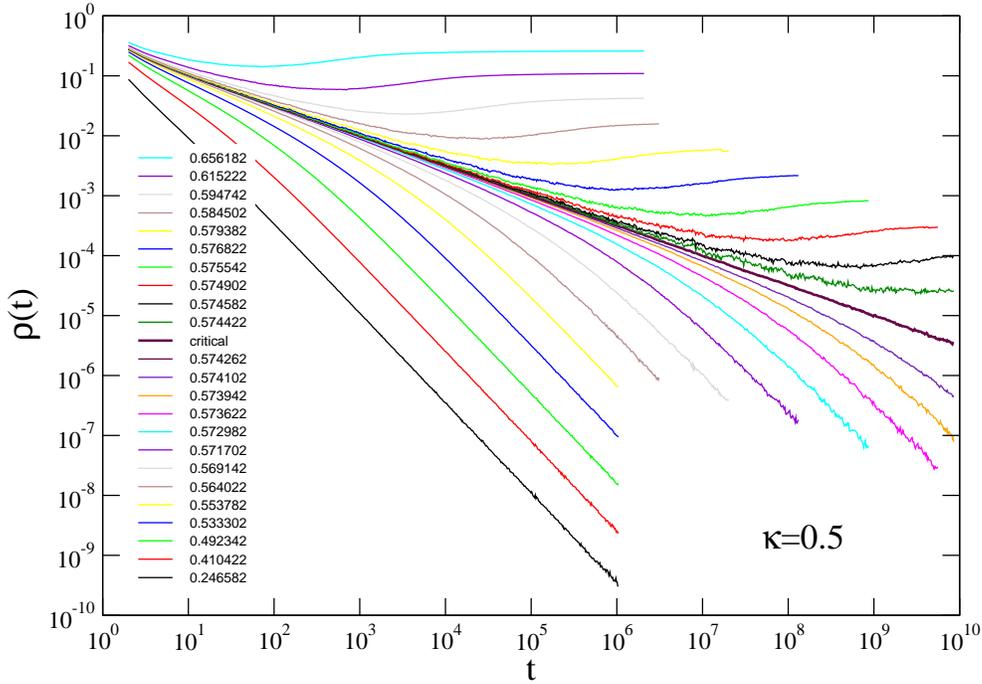}
\caption{Decay of the density of active sites in the TSP for $\kappa=0.5$ and various values of $p$, as listed in the legend. A qualitatively similar behavior is observed for all values of $0<\kappa<1$.}
\label{fig3}
\end{figure}

Starting with a single seed of activity at $t=0$ and averaging over many independent runs we measured the probability $\varrho(t)$ to find an active site at time $t$. Varying the parameter $p$ we observe the following phenomenological behavior (see Fig.~\ref{fig3}):
\begin{itemize}
\item For small values of $p$ the density of active sites decays as $\varrho(t)\sim t^{-(1+\kappa)}$. This power-law decay is a direct consequence of the L{\'e}vy distribution~(\ref{Distribution}) and characterizes the subcritical phase of the TSP.\\[-2mm]
\item For large values of $p$ the density first decreases until it reaches a minimum, then increases again until it saturates at a stationary value.\\[-2mm] 
\item At a well-defined critical threshold $p=p_c(\kappa)$ the density decays algebraically but much slower as in the inactive phase. The observed power law 
\begin{equation}
\label{densitydecay}
\varrho(t)\sim t^{-\alpha}
\end{equation}
is very clean and the critical exponent $\alpha$ is found to vary continuously with $\kappa$. It turns out that the numerical estimates are in excellent agreement with the analytical result $\alpha=1-\kappa$, as will be discussed in the following section.
\end{itemize}
For localized initial configurations such as a single seed of activity at $t=0$ the process may terminate when no active sites to be updated are left. In this case the last active site on the timeline defines the time $t_e$ where this particular run ends. 

Likewise one can estimate the probability $P_s(t)$ that the process starting with a single seed survives at least until time $t$, i.e., it produces at least one active site at time $t'\geq t$. Averaging over many runs at criticality the survival probability seems to decay algebraically as 
\begin{equation}
P_s(t)\sim t^{-\delta}
\end{equation}
although the observed scaling is less clean in this case. The survival exponent $\delta$ is significantly smaller than $\alpha$. For example, for $\kappa=0.5$ one finds $\delta = 0.165(3)$. The unusual decay of the survival propability will be discussed in details in the last section of this paper.

Continuous phase transitions into absorbing states are generically characterized by four independent critical exponents $\beta, \beta', \nu_\perp, \nu_\parallel$. The first two exponents are related to the order parameter and its conjugate field while the latter describe how the spatial and the temporal correlation lengths diverge as the critical point is approached. Interpreting the TSP as the zero-dimensional limit of directed percolation with temporal L{\'e}vy flights, it has no spatial degrees of freedom and hence the exponent $\nu_\perp$ does no longer exist. 

Moreover the question arises whether the so-called rapidity reversal symmetry, which in the case of DP forces the exponents $\beta$ and $\beta'$ to be identical~\cite{GrassbergerTorre79}, still holds for the TSP. As shown in Ref.~\cite{DeloubriereWijland02} this is indeed the case, despite the unusual value of the survival exponent $\delta$. The purpose of this study is to find out how the exponents $\alpha$, $\nu_\parallel$, and $\delta$ are related to~$\kappa$.

\section{Field theory}

In this section we describe a field-theoretic approach for the TSP which allows one to compute the critical exponents perturbatively by a loop expansion. This field theory was first introduced by Deloubri{\`e}re and van Wijland, somewhat hidden in an appendix of Ref.~\cite{DeloubriereWijland02}. Here we rederive their results independently, calculating the critical exponents to one-loop order and confirming their results.

\subsection{Langevin equation}

Before discussing the Langevin equation for the TSP, let us first to recall the well-known Langevin for \textit{ordinary} directed percolation~\cite{Janssen81} which describes the temporal evolution of the coarse-grained density of active sites $\varrho(\xvec,t)$ in a DP process in $d$ spatial dimensions:
\begin{equation}
\partial_t\varrho(\xvec,t) \;=\; a\varrho(\xvec,t) - b\varrho^2(\xvec,t) + D\nabla^2 \varrho(\xvec,t) + \xi(\xvec,t)\,.
\end{equation}
Here the first term on the r.h.s. accounts for offspring production and spontaneous removal of particles. This means that the parameter $a$ is related to the percolation probability and has to be tuned to a certain critical value at the transition. The second non-linear term prevents the density from diverging and reflects the fact that repeated activations of the same site have no effect. The third term describes nearest-neighbor diffusion in space while $\xi(\xvec,t)$ denotes a white Gaussian noise which accounts for the fluctuations of the coarse-grained density $\varrho(\xvec,t)$ caused by the stochastic nature of the dynamical rules. Since the intensity of such fluctuations depends on the density of active sites, the central limit theorem implies that the correlations of the noise are given by
\begin{equation}
\langle \xi(\xvec,t) \xi({\xvec\,}',t') \rangle \;=\; \noiseamp \varrho(\xvec,t) \delta^d (\xvec-{\xvec\,}')\delta(t-t')\,.
\end{equation}
In order to find a Langevin equation for the TSP, the DP Langevin equation given above has to be modified in two ways. On the one hand there are no spatial degrees of freedom in the present case, meaning that the spatial argument $\xvec$ as well as the diffusion term $D\nabla^2 \varrho$ have to be omitted. On the other hand, the non-Markovian dynamics by means of directed L{\'e}vy flights has to be incorporated. As shown in previous studies (see e.g.~\cite{Hinrichsen07} and references therein), this can be done by replacing the temporal derivative $\partial_t$ on the l.h.s. by a so-called fractional derivative $\tilde\partial_t^\kappa$ defined by~\cite{Fogedby94}
\begin{equation}
\tilde\partial_t^\kappa \, \varrho(t)  \;=\;  \frac{1}{\mathcal{N}_\parallel(\kappa)}
\int_0^{\infty} {\rm d}t' \, {t'}^{-1-\kappa} [\varrho(t)-\varrho(t-t')]\,,
\end{equation}
where $\kappa\in[0,1]$ is the control exponent introduced in Eq.~(\ref{Distribution}) and $\mathcal{N}_\parallel(\kappa)=-\Gamma(-\kappa)$ is a normalization constant. The effect of this operator is to transfer activity located at time $t-t'$ over a temporal distance $t'>0$ to the destination $t$ at a rate proportional to ${t'}^{-1-\kappa}$. Therefore, the fractional derivative $\tilde\partial_t^\kappa$ generates directed L{\'e}vy flights according to the distribution (\ref{Distribution}) controlled by the parameter~$\kappa$, just in the same way as an ordinary derivative $\partial_t$ generates a local translation in time. 

Technically the easiest way to handle the fractional derivative $\tilde \partial_t^\kappa$ is to consider its action in Fourier space where it brings down a factor $(-i\omega)^\kappa$ in front of the exponential:
\begin{equation}
\label{fracfourier}
\tilde \partial_t^\kappa \,e^{-i \omega t} \;=\; (-i \omega)^\kappa \,e^{-i \omega t} \,\qquad \mbox{ for } 0<\kappa<1.
\end{equation}
In this representation the directed character of the L{\'e}vy flights, which is needed to ensure causality of the temporal evolution, is reflected by the fact that $(-i\omega)^\kappa$ is not invariant under the replacement $\omega\to-\omega$. 

At first glance, the Fourier representation~(\ref{fracfourier}) suggests that for $\kappa=1$ the fractional derivative $\tilde\partial_t^1$ should act in the same way as the ordinary derivative $\partial_t$. However, it is important to note that for $\kappa=1$ Eq.~(\ref{fracfourier}) is no longer valid. In fact, $\tilde \partial_t^1$ is a non-local operator while $\partial_t$ is not. 

With these two modifications the non-Markovian Langevin equation for the TSP reads
\begin{equation}
\label{Langevin}
\tilde\partial_t^\kappa \varrho(t) = a \varrho(t) - b \varrho(t)^2 + \xi(t)\,,
\end{equation}
where $\xi$ is a density-dependent noise with the correlations 
\begin{equation}
\label{Correlations}
\langle \xi(t) \xi(t')\rangle=\noiseamp \varrho(t)\delta(t-t') \,.
\end{equation}
%

\subsection{Partition sum and field-theoretic action}

The partition sum $Z$ for the present model is defined as the sum over all configurations and all realizations of randomness which obey the dynamical rules. In the continuum limit this corresponds to the functional integration over all configurations of the field $\varrho(t)$ and all realizations of the noise $\xi(t)$ weighted according to the correlations~(\ref{Correlations}) which obey the Langevin equation~(\ref{Langevin}). Formally this may be written as
\begin{equation}
Z \propto \int D\varrho \int D\xi P[\xi] \,\, \delta\Bigl[\tau\tilde\partial_t\varrho(t)^\kappa-a\varrho(t)+b\varrho^2(t)-\xi(t)\Bigr] \,,
\end{equation}
where the Langevin equation appears as the argument of a Dirac-$\delta$ functional and
\begin{equation}
P[\xi] \propto \exp\Bigl( -\int_{-\infty}^{+\infty}\d t \frac{\xi^2(t)}{2\noiseamp \varrho(t)} \Bigr)
\end{equation}
is the functional weight of the Gaussian noise. For later convenience we introduced an additional coefficient $\tau$ in front of the fractional derivative which fixes the overall time scale of the temporal evolution.

Following standard methods described in~\cite{Janssen76a,Janssen81a} one can integrate out the noise as follows. First the $\delta$-functional is represented in Fourier space by introducing a response field $\bar\varrho(t)$:
\begin{equation*}
Z \propto \int D\varrho D\bar\varrho \int D\xi P[\xi]  \,\,\exp\Bigl[i \int_{-\infty}^{+\infty}\d t\,\bar\varrho\Bigl(\tau\tilde\partial_t\varrho^\kappa-a\varrho+b\varrho^2-\xi\Bigr)\Bigr]\,.
\end{equation*}
After a Wick rotation in the complex plane the density-dependent noise contribution can be separated:
\begin{eqnarray}
Z &\propto& \int D\varrho D\bar\varrho \,\,\exp\Bigl[- \int_{-\infty}^{+\infty}\d t\,\bar\varrho\Bigl(\tau\tilde\partial_t\varrho^\kappa-a\varrho+b\varrho^2\Bigr)\Bigr] \nonumber  \\
&& \hspace{12mm}\times \int D\xi P[\xi]  \,\,\exp\Bigl[\int_{-\infty}^{+\infty}\d t\,\bar\varrho\xi \Bigr]\,.
\end{eqnarray}
This allows the noise to be integrated, resulting into
\begin{equation}
Z \propto \int D\varrho D\bar\varrho \,\exp\Bigl[ - \int_{-\infty}^{+\infty}\d t \Bigl(
\bar\varrho \bigl[\tau\tilde\partial_t^\kappa-a\bigr]\varrho + b \bar\varrho\varrho^2-\frac{\noiseamp}{2}\bar\varrho^2\varrho \Bigr) \Bigr]\,.
\end{equation}
With $\phi(t):=\sqrt{2b/\noiseamp}\,\varrho(t)$, $\bar\phi(t):=\sqrt{\noiseamp/2b}\,\bar\varrho(t)$ and $g:=\sqrt{2b\noiseamp}$ one obtains the partition sum
\begin{equation}
\label{ActionFormal}
Z \propto \int D\phi D\bar\phi \,e^{-S[\phi,\bar\phi]}
\end{equation}
with a field-theoretic action $S=S_0+S_{\rm int}$ consisting of a free part
\begin{equation}
S_0[\phi,\bar\phi] = \int_{-\infty}^{+\infty}\d t \,\,\bar\phi(t) \biggl[\tau\tilde\partial_t^\kappa-a\biggr]\phi(t)
\end{equation}
and an interaction part
\begin{equation}
S_{\rm int}[\phi,\bar\phi] = \frac{g}{2}\int_{-\infty}^{+\infty}\d t \,\, \bar\phi(t) \biggl[\phi(t)-\bar\phi(t) \biggr] \phi(t)\,.
\end{equation}
Apart from the fractional derivative and the missing spatial degrees of freedom, this action has exactly the same structure as in the case of ordinary directed percolation. 

\subsection{Rapidity reversal symmetry}
\label{RapidityReversal}

As pointed out in~\cite{DeloubriereWijland02} the well-known rapidity reversal symmetry of directed percolation still holds in the present case, i.e. the field-theoretic action derived above is invariant under the replacement
\begin{equation}
\phi(t) \to -\bar\phi(-t)\,,\quad \bar\phi(t) \to -\phi(-t)\,.
\end{equation}
To see this it is convenient to represent the action in frequency space by introducing the Fourier transforms $\Phi(\omega):=\int_{-\infty}^{+\infty}\d t\, e^{-i \omega t} \phi(t)$ and likewise $\bar\Phi(\omega)$, turning the action into
\begin{eqnarray}
S_0 &=& \int_{-\infty}^{+\infty}\frac{\d \omega}{2\pi} \,\,\bar\Phi(-\omega) \biggl[\tau(-i \omega)^\kappa-a\biggr]\Phi(\omega)
\\ 
S_{\rm int} &=& \frac{g}{2}\int_{-\infty}^{+\infty}\frac{\d \omega_1}{2\pi}\int_{-\infty}^{+\infty}\frac{\d \omega_2}{2\pi}\,\, \bar\phi(-\omega_1) \biggl[\phi(\omega_1+\omega_2)-\bar\phi(\omega_1+\omega_2) \biggr] \phi(-\omega_2)\,,\nonumber
\end{eqnarray}
where we used Eq.~(\ref{fracfourier}). As can be seen both parts of the action are invariant under the replacement $\Phi(\omega) \to -\bar\Phi(-\omega)$ and $\bar\Phi(\omega) \to -\Phi(-\omega)$, hence the rapidity reversal symmetry still holds. This symmetry implies that the scaling dimensions of the fields $\phi$ and $\bar\phi$ have to be identical, i.e., $\beta=\beta'$, reducing the number of independent critical exponents by one.

\subsection{Dimensional analysis}

Using the rapidity reversal symmetry the theory at tree level is expected to be invariant under the scale transformation
\begin{equation}
t \to \lambda t\,,\qquad \phi\to\lambda^{-\chi}\phi\,,\qquad \bar\phi\to\lambda^{-\chi}\bar\phi
\end{equation}
with a time dilatation factor $\lambda>0$ and a field exponent $\chi=\beta/\nu_\parallel$. In the Fourier-transformed action this corresponds to the replacement
\begin{equation}
\omega \to \lambda^{-1} \omega\,,\qquad \Phi\to\lambda^{1-\chi}\Phi\,,\qquad \bar\Phi\to\lambda^{1-\chi}\bar\Phi\,.
\end{equation}
It is easy to check that this scale transformation can be compensated by changing the coefficients as
\begin{equation}
\label{CoefficientRescaling}
\tau \to \lambda^{2\chi-1+\kappa} \tau\,,\quad
a \to \lambda^{2\chi-1} a\,,\quad
g \to \lambda^{3\chi-1} g\,.
\end{equation}
In order to establish scale invariance at tree level all coefficients have to be invariant. Firstly, the invariance of $\tau$ implies that $\chi=(1-\kappa)/2$. Secondly, we have $2\chi-1=\kappa>0$ so that coefficient $a$ is relevant, hence it has to be set to zero which is just the mean-field (MF) critical point. Finally, scale invariance at tree level requires the coefficient $g$ to be irrelevant, i.e. $3\chi-1<0$ or equivalently $\kappa<1/3$. Therefore the value
\begin{equation}
\kappa_c=\frac13
\end{equation}
plays the role of a lower critical threshold where mean-field behavior sets in, comparable to the upper critical dimension $d_c$ in ordinary directed percolation. For $\kappa<\kappa_c$ the model is expected to exhibit mean-field behavior with the critical exponents
\begin{equation}
\beta^{\rm MF}=1\,,\qquad \nu_\parallel^{\rm MF}=\kappa^{-1}\,.
\end{equation}
The case $\kappa>\kappa_c$, where fluctuation effects have to be taken into account, will be addressed in the following section.

\section{Renormalization group calculation}

\subsection{Loop expansion}

In seed simulations the density $\varrho(t)$ measures the response of the system at time $t$ to an activation of a single site at $t'=0$ and therefore can be interpreted as a two-point correlation function. In the field-theoretic framework this correlation function can be expressed as
\begin{equation}
G(t-t')=\langle \phi(t) \bar\phi(t') \rangle \,,
\end{equation}
where $\langle\ldots\rangle$ denotes the statistical average according to the action. The density $\varrho(t)$ measured in seed simulations is expected to be asymptotically proportional to $G(t)$, which allows one to compute the exponent $\alpha$ in Eq.~(\ref{densitydecay}).

In order to compute the two-point function it is convenient to add external currents $J(t)$ and $\bar J(t)$ to the partition function~(\ref{ActionFormal}):
\begin{equation}
Z[J,\bar J] \propto \int D\phi D\bar\phi \,\exp\Bigl[-S_0[\phi,\bar\phi]-S_{\rm int}[\phi,\bar\phi]+\int_{-\infty}^{+\infty}\d t (J \phi+\bar J \bar \phi)\Bigr].
\end{equation}
This allows the correlation functions to be expressed as functional derivatives
\begin{equation}
G(t-t') \;\propto\; \left.\frac{\delta}{\delta J(t)}\frac{\delta}{\delta \bar J(t')}\, Z[J,\bar J]\right|_{J=\bar J=0}\,.
\end{equation}
As usual one separates the partition function into a free and an interacting part
\begin{eqnarray}
Z[J,\bar J] &\propto& \exp\Bigl(-S_{\rm int}\Bigl[\frac{\delta}{\delta J},\frac{\delta}{\delta \bar J}\Bigr]\Bigr)
\\ && \times \nonumber
\int D\phi D\bar\phi \,\exp\Bigl[-S_0[\phi,\bar\phi]+\int_{-\infty}^{+\infty}\d t (J \phi+\bar J \bar \phi)\Bigr]\,.
\end{eqnarray}
Integratinng the remaining Gaussian problem one is led to
\begin{equation}
Z[J,\bar J] \propto \exp\Bigl(-S_{\rm int}\Bigl[\frac{\delta}{\delta J},\frac{\delta}{\delta \bar J}\Bigr]\Bigr)
\;\exp\Bigl[\int_{-\infty}^{+\infty} \frac{\d \omega}{2\pi}  J(-\omega) G_0(\omega) \bar J(\omega)\Bigr]\,,
\end{equation}
where $J(\omega), \bar J(\omega)$ are the Fourier-transforms of the currents $J(t),\bar J(t)$ and 
\begin{equation}
G_0(\omega)=\frac{1}{(-i\omega)^\kappa-a}
\end{equation}
denotes the free propagator. 

At this point it is important to note that the structure of the field theory is exactly the same as in the case of ordinary directed percolation~(for a recent review see e.g. Ref. \cite{Janssen05b}). In particular, the loop expansion and the Feynman graphs are exactly the same. What changes is only the form of the free propagator and the absence of spatial degrees of freedom. Therefore, we can formally use the same loop integrals as in DP, simply omitting the integration over momenta and using the modified free propagator. For example, the one-loop expansion for the two-point correlation function is given by
\begin{equation}
G^{-1}(\omega) \;=\; G_0^{-1}(\omega) + \frac{g^2}{2}\int_{-\infty}^{+\infty}\frac{\d\omega'}{2\pi} 
G_0\Bigl(\frac{\omega}{2}+\omega'\Bigr)G_0\Bigl(\frac{\omega}{2}-\omega'\Bigr) + \mathcal O(g^4)
\end{equation}
%

\subsection{Wilsons renormalization group scheme}

For $\kappa>1/3$, where the critical behavior of the TSP is influenced by fluctuations, the integrals in the loop expansion diverge in the limit $\omega\to\infty$. In this case the bare continuum description is no longer meaningful, instead the discrete nature of the process has to be restored through the back door by regularizing the integrals. 

In the following we adopt Wilsons renormalization group (RG) scheme which turns out to be particularly suitable for the present problem. In this approach the UV divergences are regularized by introducing a cutoff in momentum space. In the TSP, where momenta are absent, we introduce a cutoff $\Omega$ in the frequencies instead, i.e., the integration range of the loop integrals is restricted to $\omega \in [-\Omega,+\Omega]$. 

Let us now consider an infinitesimal scale transformation with $\lambda=1-\epsilon$. Without a cutoff the coefficients would change according to Eq.~(\ref{CoefficientRescaling}). However, the cutoff frequency has to be rescaled as well by $\Omega \to (1+\epsilon) \Omega$, modifying the value of the integrals in the loop expansion. To take this additional change into account the integrals are evaluated within the frequency shell $|\omega|\in[\Omega,\,(1+\epsilon)\Omega]$, a process called shell integration. Finally the resulting contributions are expanded in to lowest order in absorbed in the coefficients by adding suitable terms $L_\tau$, $L_a$, and $L_g$ on the r.h.s. of the RG equation:
\begin{eqnarray}
\partial_\epsilon \ln \tau &=& 1-\kappa-2\chi - L_\tau \nonumber \\
\partial_\epsilon \ln a &=& 1-2\chi- L_a \\
\partial_\epsilon \ln g &=& 1-3\chi- L_g \nonumber
\end{eqnarray}
The loop corrections $L_\tau$, $L_a$, and $L_g$ depend on the parameters $\tau,a,g$ as well a the cutoff $\Omega$ and will be computed below.

\subsection{Renormalization hypothesis of the L{\'e}vy operator}

In preceding studies of models with long-range interactions by L{\'e}vy flights it turned out that field-theoretic loop corrections do not renormalize the fractional derivative itself, instead they always renormalize the corresponding short-range operator. Technically this can be traced back to the fact that loop expansions always yield power series in $\vec k$ and $\omega$ with \textit{integral} powers which correspond to ordinary derivatives in real space. This observation implies that fractional derivatives are modified under scale transformations exclusively by their scaling part, giving rise to an \textit{exact} scaling relation among critical exponents. Assuming the same to be true in the present case, this means that $L_\tau=0$ even in the fluctuation-dominated regime $\kappa>\kappa_c$, implying the scaling relation
\begin{equation}
\label{ScalingRelation}
\chi = \frac{\beta}{\nu_\parallel}=\frac{1-\kappa}{2}\,.
\end{equation}
This scaling relation is expected to hold exactly over the full range $0<\kappa<1$ to all orders of perturbation theory. As a direct consequence, the density in seed simulations $\varrho(t)$, which is proportional to $G(t)=\langle\phi(t)\bar\phi(0)\rangle$, is predicted to decay as
\begin{equation}
\varrho(t) \;\sim\; t^{-2\beta/\nu_\parallel} \;\sim\; t^{-(1-\kappa)}
\end{equation}
for any $0<\kappa<1$, i.e. the decay exponent in Eq.~(\ref{densitydecay}) is given by
\begin{equation}
\label{alphascaling}
\alpha=1-\kappa.
\end{equation}
In numerical simulations (see below) this prediction is confirmed with high precision.

\subsection{Analysis of the renormalization group flow}

In the second step of Wilsons RG scheme, the so-called shell integration, the propagator and the vertex coefficient change to one-loop order by
\begin{equation}
G_0^{-1}(\omega) \to G_0^{-1}(\omega) + \frac{g^2}{2} \int_{>}\frac{\d\omega'}{2\pi}G_0(\frac{\omega}{2}+\omega')G_0(\frac{\omega}{2}-\omega')
\end{equation}
and
\begin{equation}
g \to g - 2g^3 \int_{>}\frac{\d\omega}{2\pi}G_0^2(\omega)G_0(-\omega)\, .
\end{equation}
Here $G_0(\omega)=(\tau(-i\omega)^\kappa-a)^{-1}$ is the free propagator and $'>'$ denotes integration over the frequency shell $|\omega|\in[\Omega,\,(1+\epsilon)\Omega]$. Integrating and expanding to lowest order one obtains
\begin{eqnarray}
&&\int_{>}\frac{\d\omega'}{2\pi}G_0(\frac{\omega}{2}+\omega')G_0(\frac{\omega}{2}-\omega')
= \frac{\epsilon \Omega }{\pi (a^2+\tau^2\Omega^{2\kappa}-2a\tau\Omega^\kappa\cos\bigl(\frac{\pi\kappa}{2}\bigr))}+\mathcal O(\omega^2)
\nonumber \\
&&\int_{>}\frac{\d\omega'}{2\pi}G_0^2(\omega')G_0(-\omega')
= \frac{-\epsilon \Omega (a-\tau\Omega^\kappa\cos\bigl(\frac{\pi\kappa}{2}\bigr))}{\pi (a^2+\tau^2\Omega^{2\kappa}-2a\tau\Omega^\kappa\cos\bigl(\frac{\pi\kappa}{2}\bigr))}\,.
\end{eqnarray}
Therefore, the loop corrections are given by
\begin{eqnarray}
L_a &=& \frac{g^2 \Omega }{2 \pi a \left( a^2 +  \tau ^2 \Omega ^{2 \kappa } -2 a   \tau  \Omega ^{\kappa }\cos \left(\frac{ \pi \kappa }{2}\right) 
   \right)} \\
L_g &=& -\frac{2 g^2 \Omega  \left(a-\tau  \Omega ^{\kappa } \cos \left(\frac{\pi  \kappa }{2}\right)\right)}{\pi  \left(a^2+\tau ^2 \Omega ^{2 \kappa }-2 a \tau \Omega ^{\kappa } \cos \left(\frac{\pi  \kappa
   }{2}\right) \right)^2}
\end{eqnarray}
Together with the scaling relation~(\ref{ScalingRelation}) the RG equations read
\begin{eqnarray}
\partial_\epsilon \ln \tau &=& 0 \nonumber \\
\partial_\epsilon \ln a &=& \kappa - L_a \\
\partial_\epsilon \ln g &=& \frac{3\kappa-1}{2} - L_g \nonumber \,.
\end{eqnarray}
Their non-trivial fixed point is given by
\begin{equation}
L_a^*=\kappa\,, \qquad L_g^* = \frac{3\kappa-1}{2}\,
\end{equation}
or, in terms of the original paramters, by
\begin{footnotesize}\begin{eqnarray}
a^* &=&
\frac{\kappa  \Omega ^{\kappa } \left(2 (7 \kappa -1) \cos \left(\frac{\pi  \kappa }{2}\right)+\sqrt{2} \sqrt{\cos (\pi  \kappa ) (1-7 \kappa )^2+(14-17
   \kappa ) \kappa -1}\right)}{22 \kappa -2}\nonumber
\\[3mm]
(g^*)^2 &=&
\frac{8 \pi  \kappa ^2 \tau ^3 \Omega ^{3 \kappa -1} }{(11 \kappa -1)^3}
\Biggl[-\cos \left(\frac{3 \pi  \kappa }{2}\right) (1-7 \kappa )^2
\\ \nonumber &&\hspace{25mm} +\sqrt{2} \cos (\pi  \kappa ) \sqrt{\cos
   (\pi  \kappa ) (1-7 \kappa )^2+(14-17 \kappa ) \kappa -1} (1-7 \kappa )
\\ \nonumber &&\hspace{25mm} +(\kappa  (73 \kappa -22)+1) \cos \left(\frac{\pi  \kappa }{2}\right)
\\ \nonumber&&\hspace{25mm}+4 \sqrt{2}
   \kappa  \sqrt{\cos (\pi  \kappa ) (1-7 \kappa )^2+(14-17 \kappa ) \kappa -1}\Biggr]\,.
\end{eqnarray}
\end{footnotesize}
In order to determine the exponent $\nu_\parallel$ let us consider the RG flow in the vicinity of this fixed point. The linearized flow field is given by the matrix
\begin{equation}
M = \left.\left( 
\begin{array}{cc}
\partial_a a (\kappa - L_a) & \partial_a a  ( \frac{3\kappa-1}{2} - L_g)\\
\partial_g g (\kappa - L_a) & \partial_g g  ( \frac{3\kappa-1}{2} - L_g)
\end{array}
\right)\right|_{a=a^*,g=g^*}
\end{equation}
and can be computed explicitly
\begin{equation}
M = \left( 
\begin{array}{cc}
(1-\kappa)/4 & f(\kappa) \\
-2\kappa & 1-3\kappa
\end{array}
\right)\,,
\end{equation}
where
\begin{eqnarray}
f(\kappa)&=&
2 \kappa  \csc ^2\left(\frac{\pi  \kappa }{2}\right)-\frac{23 \kappa }{4}-\frac{1}{4 \kappa }+2
\\ && \nonumber 
-\frac{1}{{2 \sqrt{2}}}\sqrt{\cos (\pi  \kappa ) (1-7 \kappa )^2+(14-17 \kappa ) \kappa -1} \cot \left(\frac{\pi  \kappa
   }{2}\right) \csc \left(\frac{\pi  \kappa }{2}\right).
\end{eqnarray}
The eigenvalues of this $2\times2$ matrix can be computed explicitly. Defining the distance
\begin{equation}
\varepsilon = \kappa-\kappa_c = \kappa-\frac13 
\end{equation}
which plays a similar role as the dimensional difference $\varepsilon=d_c-d$ in field theories with spatial degrees of freedom, and expanding the eigenvalues to lowest order in $\varepsilon$ one obtains
\begin{equation}
\lambda_1=-3\varepsilon+\mathcal O(\varepsilon^2) \,,\qquad \lambda_2=\frac13+\frac\varepsilon 4+\mathcal O(\varepsilon^2)\,.
\end{equation}
The first eigenvalues is always negative while the positive one describes how the critical parameter vanishes under rescaling. Therefore, one can identify the second eigenvalue with $\nu_\parallel=\lambda_2^{-1}$, leading to the main result
\begin{equation}
\nu_\parallel = 3 -\frac94\varepsilon +\mathcal O(\varepsilon^2)\,.
\end{equation}
Together with the exact scaling relation~(\ref{ScalingRelation}) this implies
\begin{equation}
\label{beta}
\beta = 1 -\frac94\varepsilon +\mathcal O(\varepsilon^2)\,.
\end{equation}
These findings are in full agreement with the results by Deloubri{\`e}re and van Wijland in the Appendix of Ref.~\cite{DeloubriereWijland02}. Their calculation involves a dimension-like parameter $d$ which is related to the control parameter $\kappa$ in our work by $d=2-2\kappa$.

\section{Numerical results}

\begin{figure}[t]
\centering\includegraphics[width=140mm]{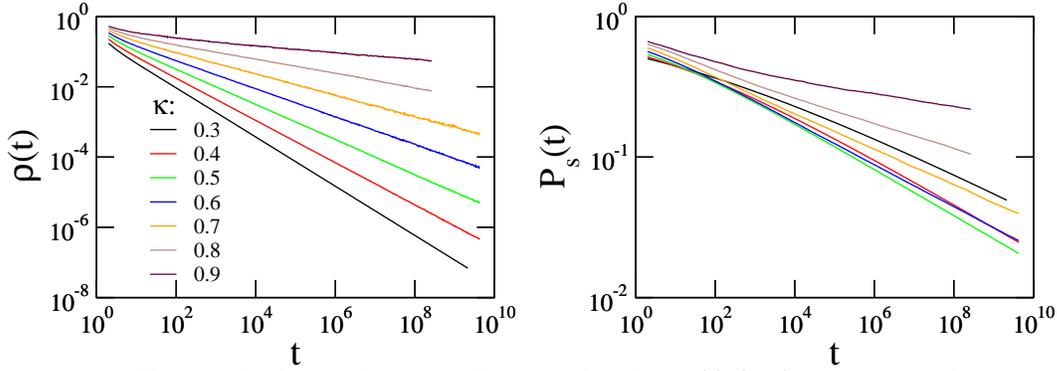}
\vspace{-5mm}
\caption{Decay of the density of active sites $\varrho(t)$ (left) and the survival probability $P_s(t)$ (right) in seed simulations of the TSP at criticality for various values of $\kappa$ averaged over $4\times 10^5$ up to $1.4 \times 10^8$ runs. The density of active sites shows a very clean power law $\varrho(t)\sim t^{-(1-\kappa)}$. The survival exponent $\delta$ is estimated by averaging over the last four decades in the right panel.}
\label{crit}
\end{figure}

We performed extensive numerical simulations with a code based on dynamical lists as described in the appendix. First we determined the critical parameter $p_c$ and estimated the exponent $\alpha$ for various values of $\kappa$. It turned out that the scaling relation $\alpha=1-\kappa$ (see Eq.~(\ref{alphascaling})) is obeyed with at least three digits accuracy in the range $0.3<\kappa<0.8$ so that we have no doubt that this scaling relation is correct. That is why we decided to consider this scaling relation as given and to use it for a precise determination of the critical point. The results are listed in Table~\ref{tab1}. As can be seen, the estimates are less accurate for very small values of $\kappa$, where the L{\'e}vy flights become extremly long-ranged, as well as in the limit $\kappa\to 1$, where the particle densities are so high so that the list-based algorithm is no longer efficient.

The same type of simulations was used to measure the survival probability $P_s(t)$ at the critical point. As shown in the right panel of Fig.~\ref{crit}, the power laws are less clean in this case. A conservative analysis leads to the estimates reported in Table~\ref{tab1}. Apart from a slight systematic deviation inside the error bars, these estimates are in good agreement with the predicted value $\delta_c$ which will be derived below in Eq.~(\ref{deltaconjecture}).

\begin{table}
\begin{center}
\hspace{5mm}
\begin{tabular}{l||l|l|l|l|l}
$\kappa$	& $p_c$ 	&$\nu_\parallel$& $\delta$ 	& $\delta_c$ & $\beta$ \\ \hline
0		& $1/2$ 	&  2		&  0		& 0 		&  1\\ 
0.1		&  0.501(1)	&  2.2(1)  	&  0.04(2)	& 0.050 	&  1\\ 
0.2		&  0.51080(2)	&  2.50(8)  	&  0.09(1)	& 0.100 	&  1\\ 
0.3		&  0.525720(5)	&  2.86(5)  	&  0.14(1)	& 0.150 	&  1\\ 
1/3 (MF)	&  0.532231(2) 	&  3.00(2)	&  0.155(10)	& 0.167 	&  1.00(2)\\ 
0.35		&  0.535762(2) 	&  2.98(2)	&  0.160(10)	& 0.169 	&  0.97(2)\\ 
0.375		&  0.541379(2) 	&  2.95(2)	&  0.164(10)	& 0.169 	&  0.93(2)\\ 
0.4		&  0.547357(2) 	&  2.92(2)	&  0.165(10)	& 0.171 	&  0.88(2)\\ 
0.45		&  0.560264(5) 	&  2.86(2)	&  0.169(3)	& 0.171 	&  0.79(2)\\ 
0.5		&  0.574262(2) 	&  2.83(2)	&  0.165(3)	& 0.167  	&  0.71(2)\\ 
0.6		&  0.604870(5) 	&  2.87(3)	&  0.148(5)	& 0.150 	&  0.57(2)\\ 
0.7		&  0.63823(1) 	&  2.98(5)	&  0.120(5)	& 0.126 	&  0.45(3)\\ 
0.8		&  0.67401(3)	&  3.17(8)	&  0.085(5)	& 0.089 	&  0.32(2)\\ 
0.9		&  0.7127(1)	&  3.9(2)	&  0.045(10)	& 0.047 	&  0.20(2)\\ 
1		&  $\approx0.76$&  $\infty$ 	&  0		& 0 		&  0
\end{tabular}
\caption{Numerical estimates of the percolation threshold $p_c$ and the critical exponents $\nu_\parallel,\delta$ for various values of $\kappa$. Entries without error bars are based on analytical arguments. The estimates for $\delta$ have to be compared with the predicted values $\delta_c$ according to the conjecture in Eq.~(\ref{deltaconjecture}) truncated to three digits. The values for $\beta$ in the last column were computed from the previous exponents using the scaling relation $\beta=\delta\nu_\parallel$.}
\label{tab1}
\end{center}
\end{table}

\begin{figure}[t]
\centering\includegraphics[width=130mm]{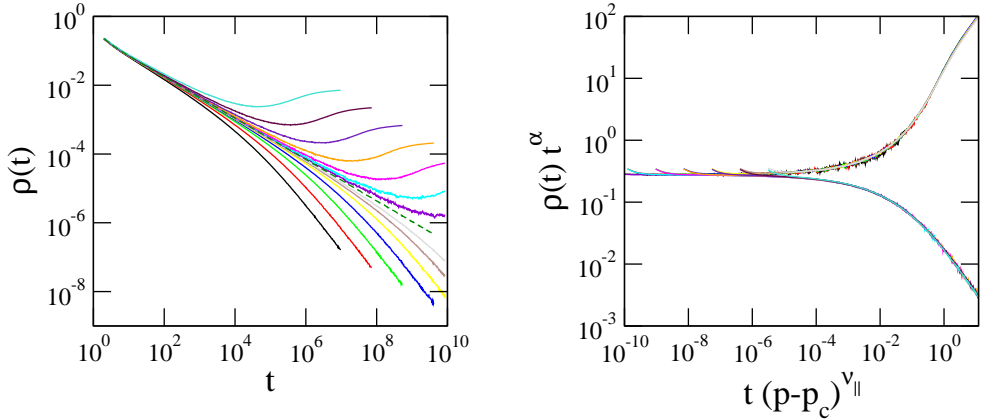}
\vspace{-5mm}
\caption{Off-critical simulations, here for $\kappa=0.4$ with $p-p_c$ ranging from $\pm 0.00016$ to $\pm 0.01024$. The measured curves (left) can be collapsed convincingly (right), allowing one to estimate the exponents $\alpha$ and $\nu_\parallel$.}
\label{off}
\end{figure}

In order to determine the exponent $\nu_\parallel$ we performed extensive off-critical simulations. A typical data set is shown in the left panel of Fig.~\ref{off}. Assuming the usual scaling form $\varrho(t,p-p_c) \simeq t^{-\alpha} R\bigl(t(p-p_c)^{\nu_\parallel}\bigr)$ and using the scaling relation $\alpha=1-\kappa$ the curves should collapse if $t^{\alpha}\varrho$ is plotted against $t(p-p_c)^{\nu_\parallel}$. As demonstrated in the right panel of Fig.~\ref{off}, one obtains a very clean collapse below and above criticality. This allowed us to estimate the exponent $\nu_\parallel$, as listed in Table~\ref{tab1}.

\begin{figure}[t]
\includegraphics[width=160mm]{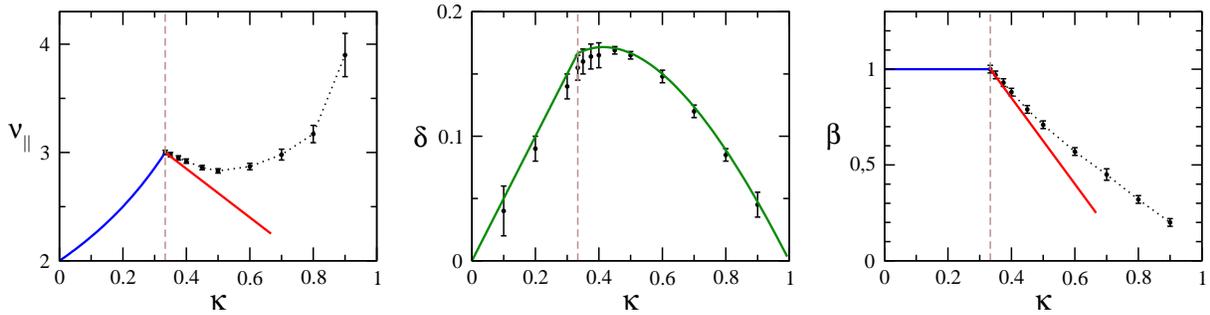}
\vspace{-5mm}
\caption{Numerically measured critical exponents (black) for various values of $\kappa$ according to Table~\ref{tab1} compared with the mean-field prediction (blue), the field-theoretic one-loop approximation (red) and the conjectured formula for $\delta$ (green), see Eq.~(\ref{deltaconjecture}). }
\label{results}
\end{figure}

As a visual summary the  estimates for the critical exponents are plotted as functions of $\kappa$ in Fig.~\ref{results} and compared with the analytical predictions. As expected, the field-theoretic one-loop expansion is tangent to the data in $\kappa=\kappa_c=1/3$, as indicated by the red lines. As can be seen, our numerical simulations fully support the field-theoretic results. They would even allow us to verify the results of a future two-loop calculation by fitting a parabola. Obviously the two-loop corrections should have the opposite sign.

\section{Related critical properties}

In this section we discuss various issues such as the conjugate field and the role of different initial conditions. Based on these arguments we arrive at a conjecture that allows us to express the survival exponent $\delta$ in terms of an exact scaling relation.

\subsection{Conjugate field}

In equilibrium critical phenomena an order parameter is always associated with a conjugate field $h$ that, when applied externally, causes a response of the order parameter. For example, in the Ising model $h$ is just an external magnetic field. The same applies to non-equilibrium phenomena as DP, where the external field corresponds to a spontaneous creation of activity at rate $h$. 

In the mean field regime of DP a constant external field causes an asymptotically stationary response $\rho_{\rm stat}\sim\sqrt{h}$. In the non-trivial regime, where fluctuation effects are relevant, the situation is different. Here a constant field $h$ is known to cause a response which scales as $h^{\beta/\sigma}$, where $\sigma=\nu_\parallel+d\nu_\perp-\beta'$. In the present case, where $d=0$ and $\beta'=\beta$, this would imply that $\sigma=\nu_\parallel-\beta$. Together with the scaling relation~(\ref{ScalingRelation}) one arrives at
\begin{equation}
\label{conjfield}
\rho_{\rm stat} \;\sim\;\left\{
\begin{array}{cc}
 h^{1/2} & \mbox{ for } \kappa\leq\frac13 \\[2mm]
 h^{\frac{1-\kappa}{1+\kappa}} & \mbox{ for } \kappa>\frac13\,.
\end{array}
\right.
\end{equation}
This power law is in good agreement with numerical results (not shown here).

\subsection{Fully occupied initial state for $t<0$}

\begin{figure}[t]
\centering\includegraphics[width=130mm]{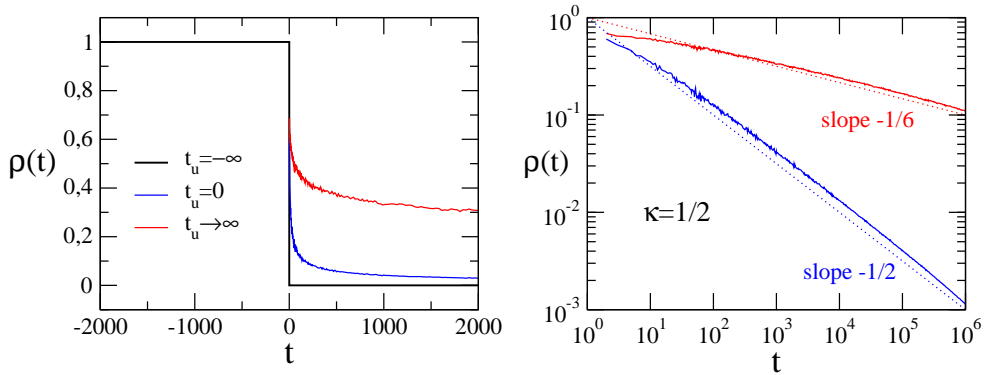}
\caption{Initial configuration where all sites at negative times are active. The left panel shows the initial configuration (black), the density profile in the moment when the update rule reaches $t_u=0$ (blue) and the final density profile in the limit $t_u \to \infty$~(red). The right panel shows the same data in a double logarithmic representation.}
\label{figfully}
\end{figure}

So far we considered simulations starting with a single active seed at $t=0$. However, as already discussed in the Introduction, the initial condition is not determined by the state of the site at $t=0$ alone, instead the non-Markovian dynamics requires us to specify the configuration of \textit{all} sites along the entire time line, in principle even including all sites at negative times $t<0$. As an example let us consider the intial configuration
\begin{equation}
\label{stepfunction}
s(t)=\left\{
\begin{array}{cc}
 1 & \mbox{ if } t\leq 0\\[2mm]
 0 & \mbox{ if } t> 0\\[2mm]
\end{array}
\right.
\end{equation}
where all sites at negative times are initially occupied. This initial density is shown as a black step function in the left panel of Fig.~\ref{figfully}. For this initial configuration the process has to start at $t_u=-\infty$. Upon reaching $t_u=0$ the update rule will have activated many sites at $t>0$. These active sites are randomly distributed and uncorrelated with an expectation value decaying asymptotically as
\begin{equation}
\varrho(t)=\langle s(t) \rangle \sim t^{-\kappa} \qquad\qquad (t_u=0)
\end{equation}
as indicated by the blue curve in Fig.~\ref{figfully}. 

Subsequently, as the update rule advances from $t=0$ to $t\to\infty$, even more sites will be activated along the timeline. The resulting density of active sites is shown as red curve in the figure. As can be seen, the density seems to decay asymptotically as $\varrho(t)\sim t^{-1/6}$ for $\kappa=0.5$. Obviously, this density is much larger than the density produced by a single seed (where one would obtain $\varrho(t)\sim t^{-1/2}$). 

\begin{figure}[b]
\centering\includegraphics[width=120mm]{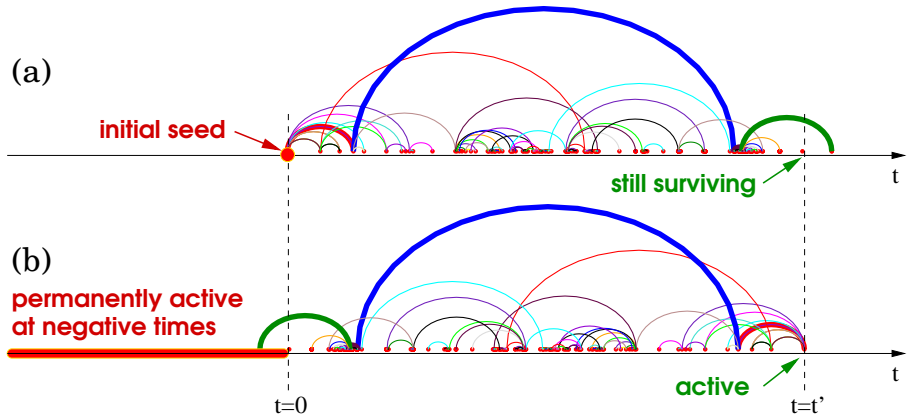}
\caption{Illustration of the time reversal symmetry. The upper part (a) shows a particular realization of L{\'e}vy flights in a simulation starting with a single seed. At time $t'$ the process is still surviving because there is at least one connected path (bold lines) that activates a site at $t\geq t'$. The lower part (b) shows the time-reversed configuration of L{\'e}vy flights which has the same statistical weight. Here the site at time $t'$ becomes active if there is at least one connected path to a site at $t\leq 0$. Therefore, a fully activated initial state at $t<0$ (indicated as a red bar) will lead to a density profile that decays in the same way as the survival probability.}
\label{figrev}
\end{figure}

Moreover, the observed exponent $1/6$ coincides with the survival exponent in seed simulations. This relationship is valid for all $\kappa$ and is a direct consequence of the time reversal symmetry discussed in Sect.~\ref{RapidityReversal}. On a qualitative level it can be explained as follows. As sketched in Fig.~\ref{figrev} a particular run in seed simulations is said to survive (at least) up to time $t$ if there is a connected path of subsequent L{\'e}vy flights from the seed to some site at $t'\geq t$. The lower part of the figure shows the same but horizontally reflected realization of L{\'e}vy flights. In such a time-reversed situation survival translates into the condition that a site at time $t$ is connected \textit{backwards} in time to some site at $t'\leq 0$, or equivalently, that a site at time $t$ is activated by an initial configuration where all sites at negative times are active. 

We therefore conclude that the survival probability in seed simulations decays asymptotically in the same way as the particle density in simulations starting with fully occupied lattice at negative times. This unusual relationship will be used in the following to determine the value of $\delta$.

\subsection{The survival exponent $\delta$}

We now propose a conjecture for an exact scaling relation that determines the value of the survival exponent. The conjecture is based on the findings of the previous subsections and works as follows. Let us consider again the TSP starting with the initial configuration~(\ref{stepfunction}) where all sites at negative times are active. As discussed before and demonstrated in Fig.~\ref{figfully} the process evolves in two steps:
\begin{enumerate}
 \item During the updates from $t_u=-\infty$ to $t_u=0$, the process produces an uncorrelated random distribution of active sites at $t>0$ which on average decays as $t^{-\kappa}$.\\[-3mm]
 \item This random distribution can be interpreted as a time-dependent external field $h(t)\sim t^{-\kappa}$, which causes the process to create additional active sites as a response during the updates from $t_u=1$ to $t_u=\infty$.
\end{enumerate}
We now assume that this external field $h(t)$ decays so slowly that the response of the process changes \textit{adiabatically} as if $h$ was stationary. Numerical experiments support the validity of this assumption over the full range of $\kappa$. This would lead to the conjecture that the final density of active sites (the blue curve in Fig.~\ref{figfully}) decays as $\rho(t) \sim h^{\beta/\sigma} \sim t^{-\kappa\beta/\sigma}$, hence the survival exponent should be given by $\delta=\kappa\beta/\sigma$. Inserting Eq.~(\ref{conjfield}) this would mean that
\begin{equation}
\label{deltaconjecture}
\delta(\kappa)=\left\{
\begin{array}{cc}
 \frac{\kappa}{2} & \mbox{ for } \kappa\leq\frac13 \\[2mm]
 \frac{\kappa-\kappa^2}{1+\kappa}  & \mbox{ for } \kappa>\frac13
\end{array}
\right.
\end{equation}
As can be seen in Table~\ref{tab1} and in the central panel of Fig.~\ref{fig3}, this prediction is in excellent agreement with the numerical estimates for the survival exponent $\delta$. This observation raises the hope that the formula~(\ref{deltaconjecture}) may qualify as an exact scaling relation.

\section{Summary}

In this paper we have investigated a spreading process on a timeline which exhibits a continuous phase transition out of equilibrium. Studying a particular model of the TSP, where active sites activate other sites in the future by means of temporal L{\'e}vy flights, we estimated the critical exponenets as functions of the L{\'e}vy exponent $\kappa$ (see Table~\ref{tab1}). Moreover, we confirmed earlier field-theoretical results obtained by Deloubri{\`e}re and van Wijland~\cite{DeloubriereWijland02} by an independent renormalization group calculation. The combination of field-theoretical and numerical methods lead to the following main results:
\begin{enumerate}
\item For $0<\kappa<1/3$ the model exhibits mean-field behavior while for $1/3\leq\kappa<1$ fluctuation effects (loop corrections) have to be taken into account. Here the lower critical value $\kappa_c=1/3$ plays a similar role as the upper critical dimension $d_c$ in other universality classes of continuous phase transitions.\\[-4mm]
\item The TSP can be interpreted as a zero-dimensional limit of directed percolation with temporal L{\'e}vy flights.\\[-4mm]
\item The phenomenological scaling theory for absorbing phase transitions is still valid for the TSP. As there are no spatial degrees of freedom, the exponent $\nu_\perp$ no longer exists, meaning that the transition is characterized by three exponents $\beta,\beta',\nu_\parallel$.\\[-4mm]
\item The time reversal symmetry, which is the essential symmetry of DP, still holds for the TSP and implies that $\beta=\beta'$.\\[-4mm]
\item As usual in such problems, the critical exponents vary continuously with $\kappa$. \\[-4mm]
\item Since the L{\'e}vy operator does not renormalize itself, one obtains the exact scaling relation $\beta/\nu_\parallel=(1-\kappa)/2$. As a consquence, in seed simualtions the density of active sites decay as $\varrho(t)\sim t^{-\alpha}$ with $\alpha=2\beta/\nu_\parallel=1-\kappa$.\\[-4mm]
\item In the fluctuation-dominated regime a field-theoretic renormalization group calculation to one-loop order leads to the approximation $\beta \approx 1 -\frac94(\kappa-1/3)$, see Eq.~(\ref{beta}).\\[-4mm]
\item Because of the non-Markovian dynamics the initial configuration requires to specify the state of \textit{all} sites, even of those at $t<0$. \\[-4mm]
\item The time reversal symmetry implies that the survival probability $P_s(t)\sim t^{-\delta}$ decays in the same way as the density of active sites generated by a process starting with a configuration where all sites at negative times are active.
\item Based on this observation we have conjectured that the survival exponent~$\delta$ is given by Eq.~(\ref{deltaconjecture}). This conjecture is in good agreement with the numerical results, see Fig.~\ref{results}. 
\end{enumerate}

\noindent
Due to the absence of spatial degrees of freedom the fiel theory for the TSP is particularly simple. It would be interesting to perform a two-loop RG calculation and to compare the results with the present numerical data.\\[3mm]

{\noindent \bf Acknowledgements}\\
We would like to thank F. van Wijland for helpful discussions. We also thank the organizers of the SigmaPhi-2008 Conference in Statistical Physics at the Orthodox Academy of Crete, were we presented the main results of this work.

\vspace{5mm}

\appendix
\section*{Appendix: Optimized code using dynamical container classes}
\addcontentsline{toc}{section}{Appendix: Optimized code using dynamical container classes}

In this appendix we demonstrate how the TSP can be simulated efficiently. For simplicity we focus on the update sequence using dynamical lists.

Most advanced programming languages provide libraries for standardized dynamical container classes such as sets, lists and maps. An example is the standard template library (STL) which became part of most C++ environements (see e.g.~\cite{Austern98a}). Since the configuration of the TSP can be characterized by an ordered set of integer numbers marking all active sites, the suitable template class to store a configuration is a so-called `set' of integers. This container class provides various functions, of which only four are important in the present case:
\begin{itemize}
 \item {\tt empty}:   returns {\tt true} if the set is empty.
 \item {\tt insert}:  adds a new element provided that it does not yet exist.
 \item {\tt begin}:   returns a pointer to the first element in the set.
 \item {\tt erase}:   removes an element from the set.
\end{itemize}
The optimized algorithm works as follows. First one includes the relevant part of the STL and creates an instance of a set of integers:
\color{darkbrown}
\begin{quote}
\begin{small}
\begin{verbatim}
#include <set>
set<long int> S;
\end{verbatim}
\end{small}
\end{quote}
\color{black}
Upon creation this set is initially empty. Note that we have used the type {\tt long int} as template argument in order to avoid  possible integer overflows caused by extremly long L{\'e}vy flights. Next, one has to specify the initial state. For a single seed at $t=0$ this can be done by
\color{darkbrown}
\begin{quote}
\begin{small}
\begin{verbatim}
S.insert(0);
\end{verbatim}
\end{small}
\end{quote}
\color{black}
Then the update loop is executed as long as the set contains elements:
\color{darkbrown}
\begin{quote}
\begin{small}
\begin{verbatim}
while (not S.empty()) tu=update(p,kappa,tmax);
\end{verbatim}
\end{small}
\end{quote}
\color{black}
In the body of this loop the function
\begin{quote}
\begin{small}
\color{darkbrown}
\begin{verbatim}
long int update (double p, double kappa, long int tmax) 
{
long int tu = *(S.begin());                    // get first element
while (rnd()<p)                                // With prob. p repeat:
   {                                           // compute target site		
   long int tnew = tu + (long int) pow(rnd(),-1/kappa);
   if (tnew<tmax) S.insert(tnew);              // activate target site;
   }
S.erase(S.begin());                            // remove first element;
return tu;                                     // return update time;
}
\end{verbatim}
\end{small}
\end{quote}
\color{black}
performs an update and returns the time $t_u$ of the updated site. In the example shown above the updated site is removed from the set in order to minimize memory consumption. Note that in a {\tt set} each element can appear only once; the attempt to insert an already existing element has no effect. In the present case this ensures that repeated activation of the same lattice site has no effect.\\


{\large\bf\noindent References: }\\[-3mm]
\bibliographystyle{revtex}
\bibliography{/home/hinrichsen/Dateien/Literatur/master}

\end{document}